\documentclass{kapproc}

\RequirePackage{graphicx}
\RequirePackage{epsf}
\input{psfig.sty}

\let\footnote\savefootnote
\let\footnotetext\savefootnotetext

\kluwerbib 

\begin{document}

\articletitle{Fifty Years of IMF Variation: The Intermediate-Mass Stars}

 \author{John Scalo}
 \affil{University of Texas, Austin Texas, USA}
 \email{scalo@astro.as.utexas.edu}

 \begin{abstract}
I track the history of star count estimates of the Milky Way field star and open 
cluster IMFs, concentrating on the neglected mass range from 1 to 15 
M${_\odot}$.  The prevalent belief that the IMF is universal appears to be  without basis for this mass range.  Two recent estimates of the field star IMF using different methods and samples give values of the average logarithmic slope $\Gamma$ between -1.7 and -2.1 in the mass range 1.1 to 4 M${_\odot}$.  Two older estimates between 2 and 15 M${_\odot}$, using much smaller samples, disagree severely; the field IMF in this range is essentially unknown from star counts. Variations in $\Gamma$ among open cluster IMFs in this same mass range have not decreased despite numerous detailed studies, with a rather uniformly distributed spread of about 1.0 to 1.4, even for studies using homogeneous data and reduction procedures and including only clusters with a significant mass range.  These cluster variations \textit{might} be accommodated by the combined effects of sampling, systematic errors, stellar evolution uncertainties, dynamical evolution, and unresolved binaries.  If so, then the cluster data are consistent with a universal 
IMF, but are also consistent with sizeable variations.  The data do not allow an estimate of an average IMF or $\Gamma$ because the average depends on the choice of weighting procedure and other effects.  There is little basis for claims that the field star IMF is steeper or flatter than the cluster IMF in this mass range.  If the spread in cluster IMFs is in excess of the effects listed above, real IMF variations must occur that do not depend much on environmental physical conditions explored so far.  The complexity of the star formation process seen both in observations and in simulations of turbulent cluster 
formation suggests that large realization-to-realization 
differences might be expected.  In this case an individual cluster IMF 
would be in part the product of evolutionary contingency in star 
formation, and the significant function of interest is the probability distribution 
of IMF parameters.  In either case an average IMF can probably be better determined from the field star IMF using data from future space astrometric missions, or from combinations of indirect constraints using integrated light or chemical evolution applied to whole galaxies.
 \end{abstract}

\section{Introduction}

    When Salpeter (1955) first derived the ``Original Mass Function," now
known as the ``Initial Mass Function" or IMF, for field stars, his primary
result was an empirical demonstration of certain fundamental aspects of
stellar evolution theory.  As it turned
out, the IMF became a fundamental datum for understanding the
formation of stars and the evolution of galaxies, and spawned diverse
techniques for its estimation, from star counts to integrated properties of
galaxies, and a bewildering variety of theories to account for its empirical form.  

The purpose of this review is to outline some of the major
developments in the field over the past decades, concentrating on recent empirical results using star counts in the mass range between about 1.2 and 15-20 M${_\odot}$ for Galactic field stars and open clusters.  It is in this range that the field star IMF can be best compared with young- to medium-age open cluster IMFs, and it is arguably the best mass range for determining a large number of cluster IMFs. The IMF in this mass range has been neglected in the past two decades as attention has turned almost entirely to
subsolar-mass stars.  While some studies have claimed some convergence on the cluster and field star subsolar IMF (see papers in this volume), there are certainly examples of differences that are so large they are difficult to understand as observational or dynamical effects (e.g. Jeffries et al. 2001 and Fig. 4a in Kroupa 2002, Fig. 2 in Kroupa \& Weidner, this volume).    Recent reviews emphasizing the IMF of subsolar stars in the field and clusters can be found in Kroupa (2002) and Chabrier (2003a), while the IMF in star-forming regions is reviewed by Meyer et al. (2000), Luhman (2002), and Hillenbrand (2004).  Useful reviews of more general IMF constraints are given by Kennicutt (1998) and Elmegreen (1999b); for earlier work see Scalo (1986).  Surprisingly, the IMF above 1 M${_\odot}$ has not been critically examined, although statements that the stellar IMF in this range is invariant are now common in the literature.  The present paper aims to examine the question of universality for this restricted but accessible mass range.

In order to focus the discussion, I omit the wealth of studies of globular clusters, the Galactic bulge, center, or halo, clusters or field stars in other galaxies, theories of the IMF, and the numerous indirect lines of evidence
on the IMF based on, for example, integrated light and chemical evolution
constraints (see other papers in this volume; Scalo 1986 for older work and more methods).
I also omit discussion of the important mass range above 15-20
M${_\odot}$ for a number of reasons.  Due to the degeneracy in mass at a given luminosity, the IMF for very massive stars 
requires spectroscopy to interpolate masses among evolutionary tracks, as
claimed by Garmany et al. (1982, see Massey 1985), and demonstrated
convincingly by Massey et al. (1995).  Unfortunately the Milky Way study of Garmany et al. (1982) has not been repeated using improved data and evolutionary tracks, and sample sizes for clusters where most of these stars are found are very small (typically a few dozen stars), giving large uncertainties, and an undersampling bias that yields an IMF that is too flat (Kroupa 2001).  Combined with the unknown sensitivity of the results to the effective temperature scale (see Massey et al. 2004), uncertainties in massive star evolutionary tracks, and details of extinction and reddening corrections for Milky Way stars (Van Buren 1985), I omit discussion of existing massive star IMF results (see Scalo 1998, Kroupa 2002 for references).  The IMF of massive stars is so important for galaxy evolution studies that a re-evaluation is urgently needed.

In what follows, ``IMF" refers to the frequency distribution of stellar masses per unit logarithmic interval of mass, dN/dlogM.  The parameter $\Gamma$ is the slope of a log-log power-law fit to some portion of this IMF, without implying that the IMF can be well-represented by a power-law or a piecewise power law; indeed other forms may give better fits to the data in some mass ranges (e.g. Chabrier 2003a). 

Salpeter's (1955) original estimate of the field star IMF was not a
power law, but had a logarithmic slope $\Gamma$ of about $-1.7$ below 1
M$_\odot$, $-1.2$ from 1 to 10 M$_\odot$, and strong steepening above that
mass (see Fig. 1).  Salpeter
suggested that if a power law were to be applied to the whole function, a
line of slope -1.35 would give an adequate representation, and this slope
became known as the ``Salpeter IMF."  Of course we know today that if the
input were updated to modern values the resulting IMF would look considerably different; in particular,
it is usually believed that the IMF is steeper at intermediate and large
masses than at subsolar masses.  Nevertheless, it has become common in the last decade to interpret various data as consistent with the Salpeter slope for masses above 1 M$_\odot$; as will be seen below, there is currently little basis for assigning a given average, let alone universal, value, at least from star counts.

Soon after the publication of Salpeter's paper, remarkable consistency of his IMF result with open cluster data was found
by Sandage (1957) using five clusters, Jaschek \& Jaschek (1957) using three  clusters, and van den Bergh (1957) using nine
clusters plus Orion, although Jaschek \& Jaschek had to assume an unrealistically large evolutionary correction (not used in the other two studies) in order to obtain agreement with Salpeter. Van den Bergh's result is shown in Fig. 1 (as the distribution per unit mass, dN/dM)
where Salpeter's original field star IMF can also be seen.  Considering the
large uncertainties in cluster IMFs that are recognized today, the
agreement was amazingly good, certainly much better than nearly any contemporary comparison of field and cluster IMFs!  

The fundamental question of IMF universality was first emphasized by van den Bergh \& Sher (1960) who, in a
pioneering study of 20 open clusters, claimed evidence for different
low-mass turnovers in the IMF. Although much of this turned out to be
incompleteness, this paper was pivotal in raising the issue of cluster IMF universality, a question that remains open today: Paraphrasing Kennicutt (1998), the answer has seemed paradoxical
because: 1. It is extremely difficult to imagine that the observed complexity of star formation results in a universal IMF; 2. The uncertainties in empirical IMF estimates may be so large that all evidence is consistent with a universal IMF.  Unfortunately this latter statement, even if true, does not help us establish whether the IMF \textit{is} universal, or, assuming it is, even on average, what that universal IMF may be.  Alternatively, real variations may contribute to the scatter in cluster IMFs.  Partly because of its appeal and simplicity (and because the physicist's need for universality runs deep), the former interpretation is nearly always adopted.  Here I emphasize the point of view that we simply do not know the answer to this question yet.

\begin{figure}[ht]
\sidebyside
{\centerline{\psfig{file=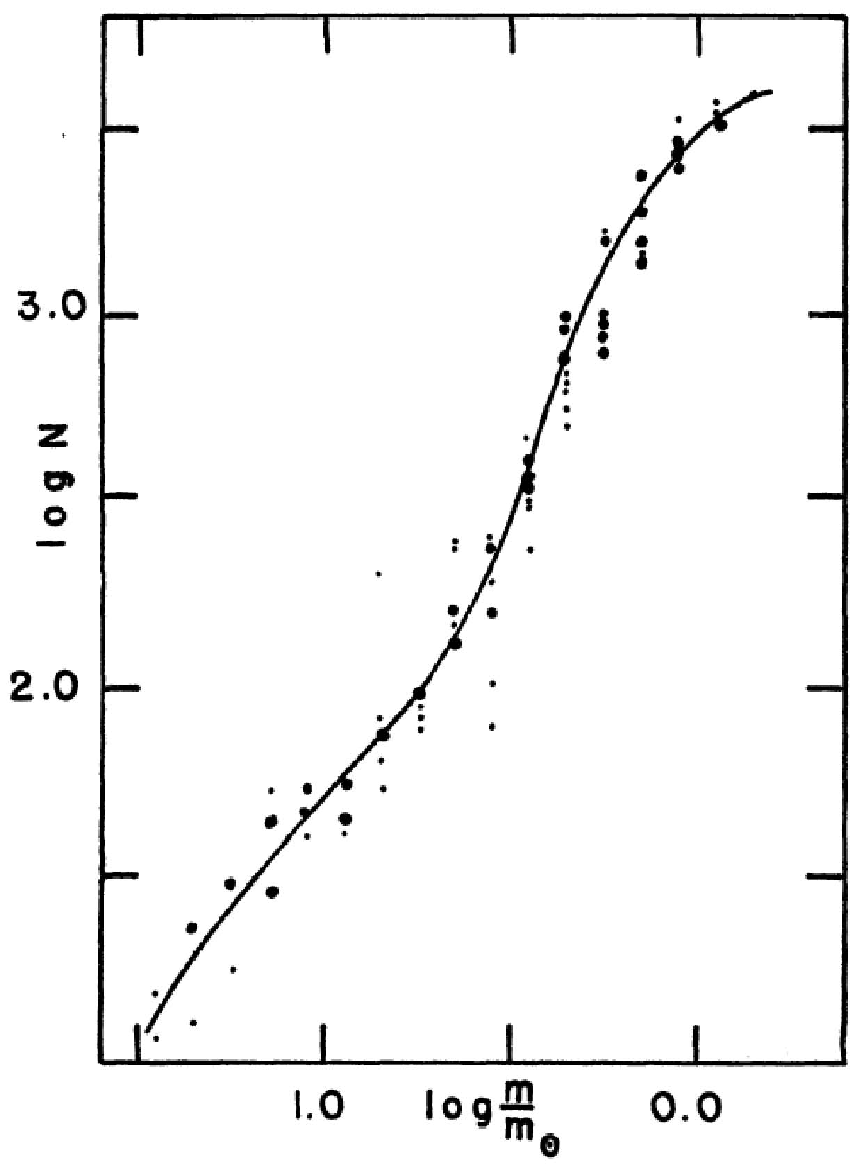,height=2in}}
\caption{IMF per unit mass combining nine clusters plus Orion compared with
Salpeter's (1955) field star IMF estimate (van den Bergh 1957).}}
{\centerline{\psfig{file=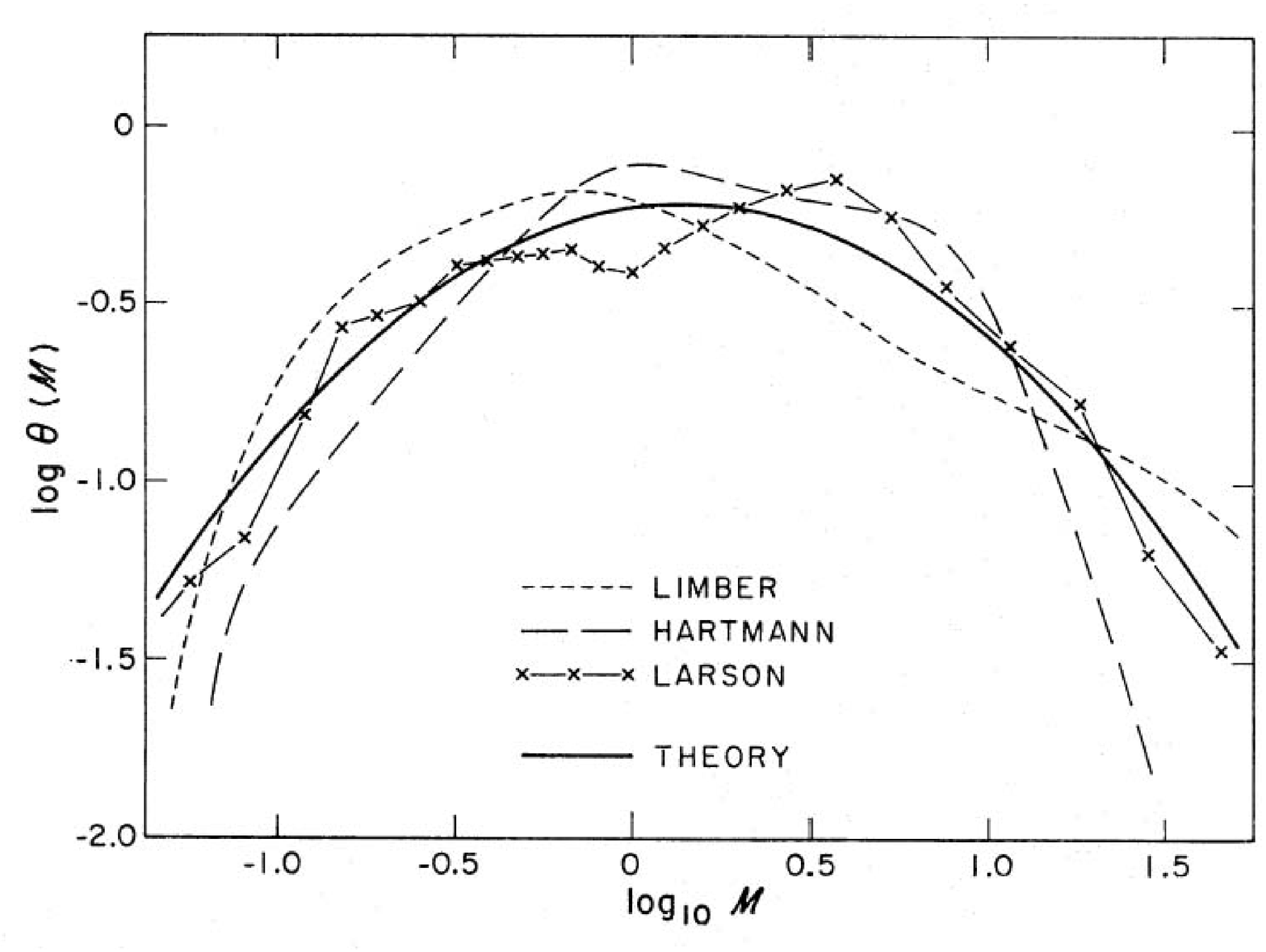,height=2in}}
\caption{Empirical estimates of the field star IMF by Limber (1960),
Hartmann (1970), and Larson (1973) as presented by Larson (1973).  The solid line is
Larson's hierarchical fragmentation model.}}
\end{figure}

\section{Historical Divergence: Field Star IMF}

By the 1970s estimates of the IMF for field stars began to
diverge from Salpeter's original estimate and from each other, a situation that has not changed much to the present day.  Fig. 2 shows three empirical
estimates for field stars compared with Larson's (1973) probabilistic
hierarchical fragmentation model.  We note that Larson was the first to
suggest a lognormal IMF.  All the field star IMFs in Fig. 2 turn over below
around 1 M${_\odot}$, but we now know that these data were very incomplete for the smaller masses.  More important is that these IMFs do not 
agree with each other even above 1 M$_\odot$, showing large
variations in shape and slope.

Motivated by this state of affairs, Miller \& Scalo (1979) reassesed the
field star IMF, which they found could be fit with half of a lognormal function, emphasizing a number of uncertain transformations and corrections to the observed luminosity function that are needed to construct the field star IMF.  A thorough analysis of more and updated observational data and theoretical input by Scalo (1986) gave a very different result.  In the mass range 1.4 to 15 M$_\odot$, where the shape of the IMF is not affected much by uncertainties in long-term SFR variations, the IMF was ``convex," rather than "concave" as in Miller \& Scalo, with a very steep
slope of $\Gamma$ = -2.6 for 1.4-3.5 M$_\odot$ and -1.8 for 2-12 M$_\odot$.
Kroupa et al. (1993) used a fit of $\Gamma$ = -1.7 to Scalo's (1986) IMF over the entire mass range above 1 M$_\odot$, but this includes the massive stars from 15 to 60 M$_\odot$, where the field star IMF was (and still is) particularly uncertain.  

Around the same time it was recognized that the construction of the mass function from the luminosity function depends very sensitively on the mass-luminosity (M-L) relation because it is the derivative of this relation that enters the conversion.  This is particularly important for the low-mass IMF (D'Antona \& Mazzitelli 1983, Kroupa et al. 1990, Kroupa et al. 1993).  Useful comparisons of empirical and theoretical M-L relations are given in Kroupa (2002) and deGrijs et al (2002a).  This development, along with the reconciliation of the long-standing discrepancy between IMFs derived from the local photometric and trigonometric parallax samples as due mostly to unresolved binaries (Kroupa et al. 1991; Chabrier 2003b) and significant advances in instrumentation, initiated a large and ongoing research effort by many groups to establish the general form of the subsolar IMF (although not well enough to decide on a functional form; see Chabrier 2003a).  Unfortunately there was no corresponding progress in the 1-15 M$_\odot$ range: the IMF derived by Yuan (1992) is essentially that of Scalo (1986), and  the ``KTG" IMF of Kroupa et al. (1993) is a joining of their new result below 1 M$_\odot$ with a single power law fit to Scalo (1986) for larger masses, and so contained no new information above 1 M$_\odot$.  

Until very recently, the only semi-independent rederivation of the IMF in the mass range of interest here was by Rana (see Rana 1991, Fig. 3) and Basu \& Rana (1992a,b).  Rana's IMF shows evidence for flat plateaus at around 2.5 and 5 M${_\odot}$, but if these features are ignored, the best fit power law IMF between 1.4 and 100 M${_\odot}$ has $\Gamma$ = -1.3, very similar to the Salpeter value.  Since the results above 15-20 M${_\odot}$, based on a (very uncertain) luminosity function, should
receive little weight for the reasons explained by Massey et al. (1995), the
corresponding single power-law fit in the 1-15 M$_\odot$ range would be significantly flatter, with $\Gamma$ around -1.  An updated version of this field star estimate was given by Basu \& Rana (1992a,b), using the
same luminosity function and mass-luminosity relation as adopted by Rana.
These papers find $\Gamma$ = -1.5 to -1.7 for masses from 1.4 M${_\odot}$ to 100 M${_\odot}$, but they note again that a single power law does not seem
to give a good fit.  Exclusion of the higher-mass stars
because of the mass degeneracy of the luminosity function would give a much 
flatter $\Gamma \approx$ -1.2.  These results are flatter by about 0.6 to 1.0 in $\Gamma$ compared to Scalo (1986) for reasons that remain unexplained.

Unfortunately there is little hope of disentangling the field star IMF from the history of the Galactic star formation rate for masses around 0.9 to 1.3 M${_\odot}$.  To make things worse, this is just the mass range over which the uncertain stellar scale height correction varies the most (Scalo 1986), and approaches the mass where main sequence lifetimes are still uncertain.  Open clusters are required to study this mass range.

Two decades of technical advances and
data accumulation, primarily Hipparcos and infrared surveys like DENIS and 2MASS, have resulted in significantly more trustworthy field star counts, at least for masses below about 4 M${_\odot}$.  Advances in interferometry, radial velocity techniques, and astrometry, have yielded many more accurate binary star masses (see Anderson 1998, Segransan et al. 2004), improving our understanding of the M-L relation.

These advances have not cleared up the situation much.  There have been only two recent determinations of the field star IMF at
masses above 1 M$_\odot$, although none that go beyond about 4 M$_\odot$.
Reid et al. (2002) combined their previous data with Hipparcos astrometry
for a 25 parsec sample of AFGKM stars in order to construct the present-day
luminosity function, and then followed the classic procedure used
by Salpeter and most subsequent work to derive the IMF.  Adopting a segmented power law to fit the results, they found IMF slopes of $\Gamma$ =
-1.8 (or -1.5) between 1.1 (or 0.6) M$_\odot$ and 3 M$_\odot$, depending on
the choice of M-L relation.  It is not clear how the
results around 1 M$_\odot$ are affected by the adopted star formation rate
history and dependence of scale height on mass.  The Reid et al. results can be compared with the steeper IMF found by Scalo (1986) and the flatter IMF found by Basu \& Rana (1992a,b) in this mass range.

An alternative method for field stars is to match synthetic populations
from evolutionary tracks to counts from Hipparcos in the color-magnitude
diagram, greatly enlarging the stellar sample and eliminating some
uncertainty due to the M-L relation and star formation history.  Schroder
\& Pagel (2003) matched counts in seven regions along the main sequence in
the HR diagram with simulated populations generated with a given IMF and SFR history.  Applying this technique to thin disk stars out to 100 pc with M$_V$ $< 4$, their best fit IMF had $\Gamma$ = -1.7 for 1.1 to 1.6 M$_\odot$, and
-2.1 for 1.6 to 4 M$_\odot$.  Schroder \& Pagel (2003) avoided many of the
considerable sources of uncertainty and bias discussed in detail by Pont \&
Eyer (2004) by matching counts in horizontal cuts through the HR diagram
rather than assigning individual masses. 

These most recent studies by Reid et al. (2002) and Schroder \& Pagel (2003) use much larger samples than earlier work, and could be optimistically interpreted as consistent with $\Gamma$ = -1.9 $\pm$0.4 in this limited mass range from 1.1 to 4 M$_\odot$, although the dependence on SFR history and scale height variation with mass is not clear.

    What about intermediate mass stars 3-4 to 15-20 M$_\odot$?  Unfortunately
determination of the luminosity function, and hence present-day mass function, 
for stars in this mass range has been eclipsed by the
availability of improved and enlarged data for lower-mass stars.  For this reason there apparently has been little improvement in the field star IMF in this mass range since the compilation of luminosity functions by Scalo (1986), which simply adopted a subjective average of the various LFs available.  A different choice of luminosity functions (and other input) in this range by Rana (1991; Basu \& Rana 1992a,b) gave a much different result.   It is probably safe to say that the field star IMF in the 3-20 M$_\odot$ mass range is unknown, or at least extremely uncertain.  Presumably future space astrometric surveys will provide trustworthy IMF estimates to at least 10-20 M$_\odot$.  For more massive stars spectroscopy and interpolation between evolutionary tracks in the HR diagram is required (Massey et al. 1995), with the attendant difficulties mentioned earlier.

\section{Historical Divergence: Open Cluster IMFs}

Studies of cluster IMFs are difficult but crucial: they offer the best
opportunity to understand the nature of the IMF and to test for
universality.  The primary advantage of clusters is of course that the
stars were all born at approximately the same time and have the same
distance and metallicity. Many of the problems mentioned above in
connection with field stars can be avoided by studying star clusters of
various ages. 

However clusters come with their own set of problems.  These
include: 1. There are few nearby clusters whose stars cover a wide range in masses, so age and limiting magnitude restricts the range of masses that can be studied in each cluster.  2. Most clusters studied have only 50-200 objects to the completeness limit, leading to a small number of stars per mass bin or a very small number of bins (assuming a histogram is used for the estimate); the associated uncertainties in estimates of $\Gamma$ are surprisingly large (Elmegreen 1999a, Kroupa 2001).  
3. Mass segregation, with more massive stars concentrated to the cluster center, occurs even in young clusters, so that it may be primordial (see Sirianni et al. 2002 and references therein), although post-formation dynamical evolution is also important (e.g. de la Fuente Marcos 1997).  Recent detailed studies of the effect in fairly young clusters can be found in Sung \& Bessell (2004, NGC 3603), de Grijs et al. (2002a, b, LMC clusters), Sirianni et al. (2002, SMC cluster NGC 330), and Littlefair et al. (2003, NGC 2547).  The ubiquity of mass segregation means that clusters must be observed to large radii.  But
then there is a severe problem with:  4. Membership: proper motions and radial velocities are needed, but this is possible only for very nearby clusters.  There are also difficult corrections that must be made for foreground and background
contamination and incompleteness.  Undercorrection for field star
contamination artificially steepens the IMF, while undercorrection for
incompleteness flattens it.  All these corrections become more important in the outer cluster regions, just the regions that must be included in order to avoid effects due to mass segregation. 5.  Unresolved binaries: this is an important effect that hides lower-mass stars in both field and cluster samples; see Sagar \& Richtler (1991), Phelps \& Janes (1993); Malkov \& Zinnecker (2001), Kroupa (2001, 2002) and references therein.
6. Adopted cluster properties: distance, age, metallicity, extinction, and differential reddening estimates can all make a difference in the derived IMF. 7. Uncertainties in evolutionary tracks and isochrones; this is especially
important for pre-main sequence stars in very young clusters (e.g. Hillenbrand \& White 2004) and very massive stars, where photometry is degenerate with respect to mass (Massey et al. 1995).  An example is shown in Kroupa (2002, Fig. 4b).  Uncertainties in the T$_{eff}$ scale enter here also.  The estimation of masses for stars from evolutionary tracks may be subject to the same types of insidious and large uncertainties that plague the estimation of ages of field stars near the main sequence by interpolation in isochrones, and need to be addressed in detail using an approach along the lines of Pont \& Eyer (2004).  8. Cluster IMFs could also be affected by preferential escape of low-mass stars during dispersal of residual gas in initially virialized clusters (Kroupa \& Boily 2002).

Many of these effects increase with cluster distance.  It is
important to notice that mass segregation, unresolved binaries, and ejection during residual gas dispersal all tend to give a cluster IMF that is too flat relative to the true or original cluster IMF; for that reason \textit{every estimate of $\Gamma$  derived for clusters should be considered an upper limit} (i.e. the IMF could be significantly steeper).

There were many numerous post-1960s estimates of IMFs in open clusters.  Most of this early work was reviewed and reproduced using a uniform mass-luminosity relation in Scalo (1986).  Some highlights include: 1. Taff's (1974) composite IMF for 62 clusters that could be well-fit by a single power law of slope -1.7 from 1 M$_\odot$ to at least 20 M$_\odot$; unfortunately Taff did not comment on the variations among clusters, and his procedure for combining cluster data has never been re-examined. 2. The IMFs for five young clusters and OB associations using evolutionary tracks combined with Stromgren photometry in the mass range 2.2 to 10 M$_\odot$ by Claudius \& Grosbel (1980) was a model for future studies of young clusters.  Their  $\Gamma$ values were estimated using a maximum likelihood method rather than least squares fitting to histograms.  The results gave $\Gamma$ from -1.6 to -2.0, but most of the samples were small; combining the Orion subgroups using 135 stars gave $\Gamma$ = -1.9, similar to that derived independently by Brown et al. (1994).  
3.  A study of 75 clusters with a range of ages by Tarrab
(1982) found logarithmic slopes that varied enormously, from near zero to
about -3, although most were in the range -1 to -2.  Her study necessarily
used inhomogeneous and old sources of data, many of the clusters only have about
20-50 members used in the IMF determination, and a few of these clusters have been studied more recently with different results (although Sanner \&
Geffert 2001 find good agreement for four of five clusters in common).

The spread in $\Gamma$ found by Tarrab (1982) has, surprisingly, not decreased much in later comparisons of clusters (see Fig. 5 in Scalo 1998, reproduced with a few additions above 1 M$_\odot$ in Kroupa 2002; also Kroupa \& Weidner, this volume; and more recent Milky Way cluster studies summarized below).  According to Tarrab, only a small part of the scatter could be attributed to sampling fluctuations.  However a careful study of this problem by Kroupa (2001) suggests that much of the spread can be due to a combination of sampling noise (see Kroupa's Fig. 3), dynamical effects, and (for low-mass stars) unresolved binaries.  The question of how much of the spread in IMF slopes among clusters is real remains central question today, and is discussed further below.

Since the 1980s there have been significant instrumental, observational, and
theoretical advances that have opened up new ground for cluster IMF studies.
CCD arrays on large telescopes allowed accurate, deep photometry. Multi-object spectrographs allowed IMF
determinations in very young clusters where interpolation between isochrones
is necessary (e.g. Luhman 2002 for a review). Sensitive near-IR cameras
allowed the study of IMFs of embedded star clusters using near-IR luminosity functions, an approach that has been greatly refined in the past few years (see Lada \& Lada 2003, Muench et al. 2003), especially in combination
with spectroscopic surveys (e.g. Hildebrand \& Carpenter 2000, Luhman et al.
2003, Wilking et al. 2004).  These efforts concentrate on subsolar- and substellar-mass stars.  New generations of stellar evolutionary models use improved input physics, cover a broad range of masses and metallicities, include pre-main sequence and brown dwarf evolutionary tracks, and are often publicly available.  However the theoretical models and the empirical effective
temperature scale still have considerable uncertainties, even for main sequence stars (e.g. Hillenbrand  \&  White 2004, Pont  \&  Eyer 2004).

What do Milky Way open cluster studies using these improved techniques and models tell us?  I will concentrate on masses above 1 M$_\odot$ and the
issue of universality, and except for a few cases only cite papers later than 1998; most earlier work is discussed at length in Scalo (1998). In nearly all the cases listed below the number of objects used is of order 100 or larger and the quoted fitting uncertainties are less than $\pm$0.25.

One of the best examples of a cluster-to-cluster IMF variation
remains the two clusters NGC 663 and NGC 581, part of a study
of eight youngish clusters by Phelps \& Janes (1993).  Both clusters have
similar ages, number of members, and limiting magnitude, the raw data were
obtained and reduced by the same authors in the same way, and both show
exceptionally fine power-law forms for their IMFs, yet $\Gamma$ = -1.1 for
NGC 663 while $\Gamma$ = -1.8 for NGC 581, each with very small fitting uncertainties.

An increasing fraction of studies find a steep $\Gamma$ around -2 in the mass range (1-2) to (4-10)
M$_\odot$.   These include NGC 2422 (Pisinzano et al. 2003, corrected for dynamical evolution), Stock 2 (Sanner \& Geffert 2001), NGC 2323 (Kalirai et al. 2003), NGC 4815 (Prisinzano et al. 2001), NGC 6631 (Sagar et al. 2001), the Pleiades (Sanner \& Geffert 2001; see also
Moraux et al. 2004 and the comparisons in Prizinzano et al. 2003 and Kroupa 2002; Kroupa \& Weidner, this volume, Fig. 2), and the Orion Nebula Cluster for masses above
about 2.5 M$_\odot$ in the spectroscopic survey of Hillenbrand (1997, Hillenbrand  \&  Carpenter 2000) when stars out to 2.5 pc are included.

Examples of much flatter IMFs with $\Gamma \approx$ -0.9 in similar
mass ranges include: NGC 2244 (Park \& Sung 2004), NGC 2451A (Sanner et al. 2001, although the sample size is very small), the starburst cluster NGC 3603 (Sung \& Bessell
2004), and several earlier studies listed in Scalo (1998).  Some of these flat IMFs could be due to mass segregation, but some of these studies tried to include those effects in their estimates.  Of course there are
examples everywhere in between, including many that are around -1.1 to -1.6, so could be consistent with the Salpeter value considering uncertainties;
e.g. Sanner et al. (2000) for NGC 1960 and NGC 2194, Slesnick et al. (2002) for h and $\chi$ Persei, Yadav \& Sagar (2004)
for NGC 2421, Yadav \& Sagar (2002) for Tr 1 and Be 11.  One of the best cases appears to be NGC 7510 with $\Gamma$ = -1.1 for 1 to 13 M$_\odot$ (Sagar \& Griffiths 1998, see Fig. 4 below).  The
comprehensive study of the Upper Scorpius OB association by Preibisch et al.
(2002) gave $\Gamma\approx$ -1.6 between 2 and 20 M$_\odot$.

Perhaps the spread is due to different reduction procedures, M-L
conversions, evolutionary tracks, even binning procedures, etc. used by different groups, as suggested by Massey (2003).  If we look
at studies that compare a number of clusters using homogenous
data and the same reduction procedures, the degree of variation seems just as large.  In
various mass ranges spanning 1 to 15 M$_\odot$, the ranges in $\Gamma$ for
some such programs come out -1.1 to -1.8 (Phelps \& Janes 1993, seven
clusters); -0.7 to -2.3 (Sanner \&
Geffert 2001, seven clusters,
although sample size is small in all but three cases; the IMFs for six of
these clusters are reproduced in Fig. 3; see their Fig. 4 for the Pleiades); and -0.8 to -1.9 (De Marchi et al., this volume, including only the five clusters with sufficient data above 1 M$_\odot$).  Using a sample of six very young but  distant clusters (with correspondingly large uncertainties), Piskunov et al. (2004) obtained a range of $\Gamma$ from -1.0 to -1.5, suggesting consistency with a single $\Gamma$ close to -1.3, given the very large quoted fitting uncertainties.  However Sagar and coworkers (see Sagar 2002 for a review) have studied a large number of distant clusters, and found an extremely large range in $\Gamma$, even including only the clusters that have IMF fitting errors of less than about $\pm$0.25, from -0.3 to -2.5 (e.g. Sagar \& Griffiths 1998, Subramaniam \& Sagar 1999,
Sagar, Munari, \& de Boer 2001), although
they tend to cluster around -0.8 to -1.8, and the authors summarize their
results as being consistent with the Salpeter value for the mean, implying that the actual uncertainties are very large.  The IMFs for the five clusters studied by Sagar \& Griffiths (1998) are shown in Fig. 4.  

It appears that the range in $\Gamma$ among clusters is not decreased when we only consider homogeneous open cluster studies.

\begin{figure}[ht]
\sidebyside
{\centerline{\psfig{file=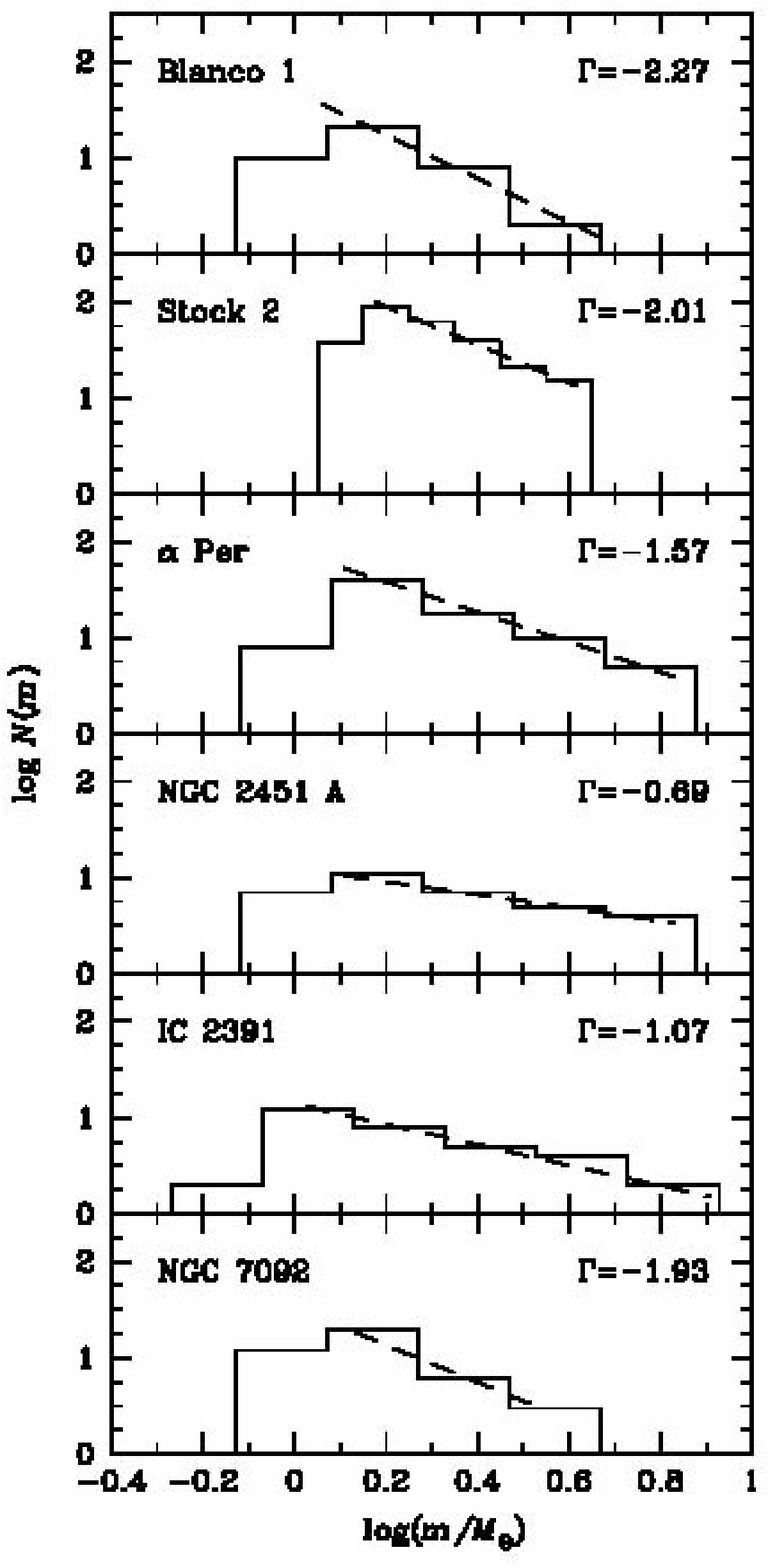,height=3in}}
\caption{IMFs for six clusters studied by Sanner \& Geffert (2001).}}
{\centerline{\psfig{file=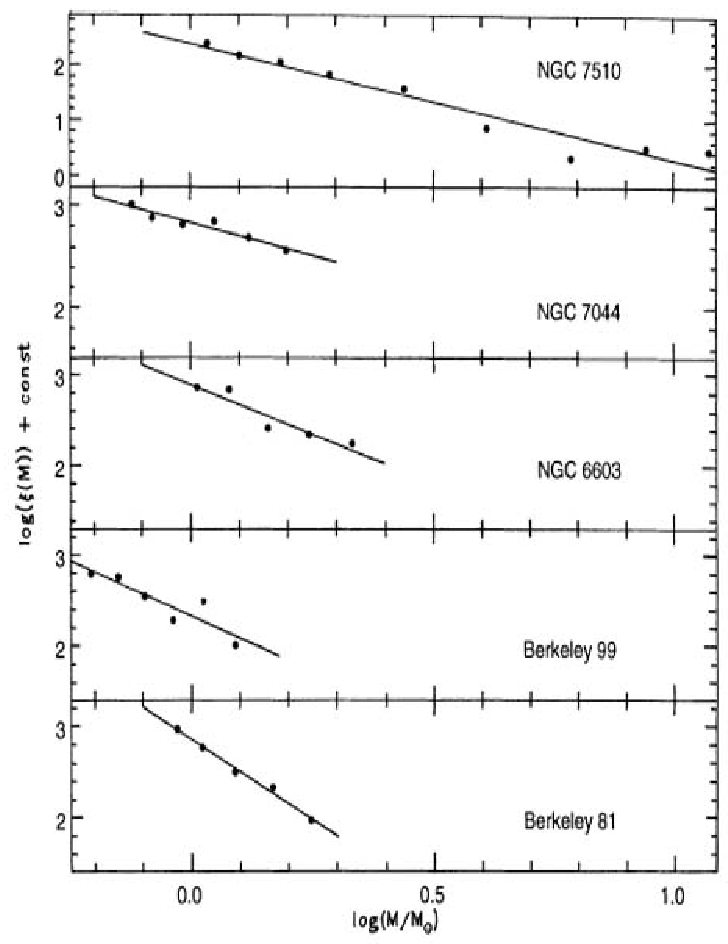,height=3in}}
\caption{IMFs for five clusters studied by Sagar \& Griffiths (1998).  The older clusters are dynamically evolved but have similar richness and ages $\sim$ 1 Gyr.}}
\end{figure}

Porras et al. (2003 and references therein) have nearly completed the
J-band luminosity-function modeling of 40 young embedded clusters along the
Perseus Arm.  Although there is considerable spread in the individual $\Gamma$ values (the individual IMFs are fairly noisy), perhaps the most interesting result is that most of the cluster IMFs, which extend considerably above 1 M$_\odot$, give a mean value of $\Gamma$  = -1.0 $\pm$0.14, significantly flatter than other studies discussed above.

There are many other intriguing cases of IMF variations in clusters that are not
easy to dismiss, including low-mass cutoffs well
above the limiting magnitude of the study, e.g. the 3 M$_\odot$ turnover in NGC 6231 (Sung et al. 1998) and "dips" at certain masses that are difficult to ascribe to random sampling effects (NGC 3293, Baume et al. 2003; NGC 6231, Sung et al. 1998; NGC 2571, Georgi et al. 2002; NGC 2580, Baume et al. 2004; and the Orion Nebula Cluster, see Hillenbrand \& Carpenter 2000). There may even be an example of a cluster with different IMFs in different subregions (NGC 7654, Pandey et al. 2001).

Altogether the spread in these studies is about 1.4 in $\Gamma$.  It is difficult to ascribe this spread primarily to sampling noise, as
proposed by Kroupa (2001), since there is no discernible correlation between
the departure from the straight mean cluster value and the sample size (Scalo, in preparation).  Strong \textit {differential} effects of dynamical evolution, even among clusters with similar richness and age, could be important, since the areal coverage (say relative to the half-mass radius) of different clusters varies significantly.  Differential unresolved binary effects would also need to be invoked, which should give an apparent flattening with distance.

The standard approach in the past decade has been to ascribe all variation to noise, errors, and systematic effects, and take a straight mean of the cluster $\Gamma$ values as representative of a universal IMF (see Kroupa 2002).  When carried out for the enlarged sample described here, the average $\Gamma$ is about Ð1.6 in the range 2 to 15 M$_\odot$, but there are various problems with such a procedure.  The most severe problem is how the various
determinations of $\Gamma$ in clusters should be weighted or even included in taking an
average; for example, if weights are assigned according to total sample size or
average number of stars per mass bin, or cluster distance, the resulting averages differ from the straight unweighted average.  An extreme example is: should we include the Perseus arm clusters with flat IMFs studied by Porras et al. as 38 points with low weight, or a single point?

In principle it is possible to estimate the pdf of the cluster $\Gamma$ values to see if it peaks at some value that could be regarded as the most likely, or average (Kroupa 2002).  This pdf, constructed by combining the Galactic sample of clusters in Scalo (1998) and enlarging it to include more recent studies (many of which were cited above), does not resemble a Gaussian as might be expected for pure sampling errors when only clusters with mean (log) masses from 1.5 to 15 M$_\odot$ and clusters younger than about 0.5 Gyr (to avoid main sequence turnoff problems) are included.  However such a pdf estimate is jeopardized by the almost certain need for segregation and unresolved binary corrections, and especially the subjective choice of studies to be included.  That the pdf above about 2 M$_\odot$ is consistent with a Gaussian (Kroupa 2002) is not supported for this enlarged sample; Kroupa's (2002, Fig. 5b, also Kroupa \& Weidner, this volume) sample in this mass range is identical with Scalo (1998) except for the addition of two more recent Milky Way studies, and was influenced by the inclusion of several studies of massive star IMFs both in the Milky Way and Magellanic Clouds, which were omitted from the present sample.  Nevertheless, the pdf of $\Gamma$ values as used by Kroupa (2002) provides an important approach to the universality problem, and needs to be reconsidered in detail.

Even if the IMF varies from cluster to cluster, an average IMF might
be meaningful, and it is of interest to ask whether this average, if
universal,  is the same as the field star IMF.  It was recognized by Reddish (1978) and Vanbeveren (1982, 1983)
that a potential problem in such a comparison is that if the upper stellar
mass limit in \textit{most} clusters is limited by the available cluster mass, the field star IMF will be too steep compared to the "true" cluster IMF
because the mass distribution of clusters gives a smaller probability for
obtaining more massive stars. 
Kroupa \& Weidner (2003, also this volume)
independently examined this problem in more detail, concluding that there should be a significant difference between the field and cluster IMF.  However the effect depends on the assumed lower limit of cluster masses; the field-cluster IMF difference basically sets in at the lower limit assumed for the cluster mass function (see the case chosen in Goodwin \& Pagel 2004).  Empirically it is not at all clear that the field and average cluster IMFs differ, or in what direction.  This is an important question, since the effect could significantly alter the appropriate IMFs for galaxy evolution studies (Weidner \& Kroupa 2004).

\section{Conclusions}

The surprisingly good agreement between the field star and cluster IMF found in the late 1950s had changed to divergence by the 1970s.  The main point of this review is that re-convergence has not yet occurred in the mass range 1 to 15 M$_\odot$, despite claims to the contrary.  Different determinations of the field star IMF above 1-2 M$_\odot$ have continued to disagree, and the field star IMF above about 4 M$_\odot$ has not been throroughly reinvestigated in nearly two decades, despite large differences between existing studies.  The differences in $\Gamma$ between open cluster IMFs found in the 1980s have not decreased despite advances on several fronts; values of $\Gamma$ appear rather uniformly distributed with a spread of around 1.0 to 1.4, even for groups of clusters with about the same age studied using the same reduction procedures.  

These cluster variations might still be accommodated by the combined effects of sampling, systematic errors, dynamical evolution, and unresolved binaries, as suggested by Kroupa (2001; also Kroupa \& Boily 2002).  In that case the necessary importance of the latter two effects implies that the cluster mass functions are steeper than determined, and that the observations are only consistent with a universal IMF because these uncertainties are so large; the observations do \textit{not} imply that the IMF \textit{is} universal.  In fact, given the evidence and problems discussed earlier, it should be clear that just about any hypothesis concerning the field star and cluster IMFs could be accommodated in the 1-15 M$_\odot$ mass range.  Appeals to Occam's Razor appear forced in this case. A balance must be struck between convenient interpretations of disparate results, and attention to uncertainties, physical effects, and convincing evidence, mirroring the cautions discussed by Brandl \& Anderson (this volume) in the context of starbust IMF estimates.

In spite of these results, the prevalent notion about the IMF remains that the IMF is approximately universal.  A key reason for this confluence of opinion, besides its simplicity, is that there is no evidence for a \textit{systematic} dependence of the IMF on gas or star density, metallicity, galactocentric radius, or time in our own or any other galaxy (with the possible exception of starburst galaxies and the Galactic center; but see Figer 2004, Brandl \& Anderson 2004, Leitherer 2004, this volume), a point that has been made for both the high-mass and low-mass IMF by many authors (e.g. Massey \& Hunter 1998; see Luhman 2002, Kroupa 2002 for more discussion).  Since these are the types of variations that astrophysicists might have expected, the default position became universality, with a corresponding projection of this belief onto interpretation of the observations. 

However, complex physical processes can exhibit variations in their statistical properties not coupled primarily to average environmental conditions.  Variations can be due to sensitivity to initial conditions, or contingency in complex evolutionary processes, as suggested for massive stars by Bonnell et al. (2004).  If the high-mass IMF is stochastic, and is coupled to the formation of the rest of the IMF, e.g. through disk photoevaporation or other feedback (see Silk, this volume), the whole IMF would be expected to show large ``tail wagging the dog" variations (see Robberto et al., this volume).  Even without such feedback, the final mass of a gas clump destined to become one or more stars is a chancy matter in a turbulent star-forming region.  It will be interesting to investigate the level of IMF variations between realizations of MHD turbulence simulations when such calculations  have the resolution and physical realism to actually the formation of individual stars.  See Jappsen et al. (2004; Klessen et al., this volume) for an example.  In this case the significant empirical function of interest is the probability distribution of individual IMF parameters, as recognized by Kroupa (2002).  However if the average IMF is the quantity of interest, field star results from future astrometric missions and careful combinations of integrated light (see Kennicutt 1998; Leitherer, this volume) and chemical evolution constraints (e.g. Renzini, this volume) may be less problematic.

\end{document}